\documentclass{iopart}

\usepackage[T1]{fontenc}
\usepackage{ae,aecompl}

\usepackage[latin9]{inputenc}
\usepackage{graphicx}
\usepackage{amssymb}

\begin{document}

\title{Quantum Analogies in Ionic Transport Through Nanopores}

\author{Andrew Meyertholen}


\address{Department of Physics, University of
Toronto, 60 St. George Sreet, Toronto, Ontario, M5S 1A7, Canada}
\ead{ameyerth@physics.utoronto.ca}

\author{Massimiliano Di Ventra}

\address{Department of Physics, University of California, San Diego, California 92093-0319, USA}
\ead{diventra@physics.ucsd.edu}

\begin{abstract}

Ionic transport in nanopores or nanochannels is key to many cellular processes and is now being explored as a method for DNA/polymer sequencing and detection.
Although apparently simple in its scope, the study of ionic dynamics in confined
 geometries such as nanopores - when the microscopic details of the surrounding environment are properly taken into account - has revealed interesting new phenomena that have an almost one-to-one correspondence with the quantum regime. The picture that emerges is that ions can form two `quasi-particle' states, one in which they surround themselves with other ions of opposite charge - {\it ionic atmosphere} - and one in which semi-bound water molecules form layers at
 different distances from the ions - {\it hydration layers}. While the first quasi-particle state has less relevance in experiments of ionic flow in nanochannels that are presently pursued, the second state gives
 rise to two additional effects. In the first, which is a {\it single} quasi-particle effect, the ionic conductance through a nanopore of given radius is predicted to be `quantized' as a function of pore radius, with the corresponding `quantization units' not
 related to universal constants - like the Plank constant $h$ and the elementary charge $e$ -, but rather to the radii of the hydration layers. The second effect instead involves the {\it many-body} interaction
  among ionic quasi-particles of the same sign, and occurs when the pore has a finite capacitance to accommodate ions so that there is a threshold concentration beyond which ions of the same sign are
  not energetically allowed to enter the pore. This effect is the equivalent of the Coulomb blockade effect one encounters in mesoscopic and nanoscopic systems of finite capacitance set out of equilibrium.
  Like the same effect in the electron transport case, the ionic counterpart appears only in the `quantum' regime, namely when the hydration layers forming the ionic quasi-particles need to break in order to pass
  through at least one of the openings of the pore. Here, we review all these phenomena, and discuss the conditions under which they may be detected. Along the way, we make the analogy with the electronic quantum
  transport case, pointing out both the similarities and differences. Since nanopores are being considered for a host of technological applications in DNA sequencing and detection, we expect these phenomena will become
  very much relevant in this field and their understanding paramount to progress.

\end{abstract}

\maketitle

\tableofcontents

\section{Introduction}
\label{Introduction}
Many crucial cellular processes involve ionic transport through nano-sized channels, so that ion dynamics is a ubiquitous effect in living organisms~\cite{Hille_01, Unwin_89, Parsegian_69, Blaustein_04}.  However, in the past decade or so, ionic transport in biological and/or synthetic nanopores (or nanochannels) has been garnering considerable new attention in view of the many applications it offers in DNA/polymer detection and sequencing~\cite{Zwolak_08,Branton_08}. In this case, nanopores hold the promise of cutting costs and reducing sequencing time significantly, thus opening the possibility for many exciting advancements, most notably ``precision medicine'': the ability to targeting a drug to a specific genome, rather than to an average population~\cite{Zwolak_08,Branton_08}.

For instance, initial work focused on ionic transport through a nanopore immersed between two ionic reservoirs~\cite{Kasianowicz_96,Akeson_99}. DNA strands translocate through the pore in response to an external bias.  This bias creates a background ionic current which is blockaded when the DNA obstructs the pore entrance.  Early experimental attempts of this procedure by Kasianowicz {\it et al.}~\cite{Kasianowicz_96} indeed measured a blockade current associated with the translocation of DNA strands.  Since then much progress has been made toward the eventual goal of single nucleotide resolution during DNA translocation.  However, the size and nature of the nanopores pose serious limitations to single-nucleotides differentiation~\cite{Stoddart_09,Clarke_09,Ashkenasy_05,Meller_01}: too many nucleotides typically reside in the pore, making the resolution of individual nucleotides difficult. Recently, individual detection has been possible via the use of exonuclease to cleave the strands, thus sending nucleotides though the pore one at a time~\cite{Stoddart_09}.  This method employs the use of ``designer'' pores, which are pores modified with molecular adapters that aid in creating a more easily differentiated blockade current.

Another proposal for DNA sequencing and detection has focused on measuring the transverse tunneling currents of each nucleotide in the pore as translocation occurs~\cite{Zwolak_05}.  With the insertion of probes in the nanochannel one may measure the distribution of electrical characteristics of the nucleotides~\cite{Lagerqvist_07,Dubi_09,Zwolak_09,Krems_09, Krems_10,Zwolak_10,Rosa_12,Krems_13,Doi_13}.  Similar to a scanning tunneling microscope, these probes may be on the sub-nanometer range, and due to the sensitivity of tunneling currents, this approach may indeed differentiate between the four DNA bases.  In fact, theoretical work combining molecular dynamics simulations with
quantum transport calculations has shown all four nucleotides to be statistically distinguishable via tunneling~\cite{Lagerqvist_06}. These results have been recently confirmed experimentally thus lending support to this sequencing idea~\cite{Tsutsui_10,Chang_10,Kawai2012}.

However, in all these cases - and any other case in which nanopores are envisioned as tools for DNA interrogation - a better understanding of ionic transport in restricted geometries is necessary. In the first proposal, for instance, ions play the major role in the differentiation of nucleobases. In the second, ion
dynamics needs to be understood in order to differentiate its contribution from that of ordinary electrical currents.

The problem, however, is far from trivial. Both biological and synthetic nanopores are operated in aqueous environments, leading to complex interactions between ions, pore and water. What has emerged recently is that
this complex environment leads to phenomena that have a quantum analog in the electrical domain~\cite{Zwolak_09, Zwolak_10,Krems_13}. Like in the quantum case, these phenomena originate fundamentally from ionic ``quasi-particle" states, which are known to form in a liquid environment~\cite{Wright_07} thus making this analogy even stronger. The correspondence between this classical type of phenomena and their quantum counterparts is the topic of this review.

Even though the concept of ionic "quasi-particles" - such as ionic atmosphere~\cite{Wright_07} and hydration layers~\cite{Hille_01} - has been known for quite some time, their fundamental role in ionic transport phenomena in nanopores such as the ones we will discuss in this review has not been previously stressed. In addition, many previous theoretical studies have focused on models which treat the nanochannel as a continuum~\cite{Kamenev_00, Zhang_05}.  In this picture, the channel is treated as an electrostatically one-dimensional system, since the dielectric constant of water is much larger than that of the surrounding pore material.  This leads to a large potential energy barrier and therefore to an exponentially large channel resistance. Additionally, it was shown that the existence of even a small amount of surface charges reduces this energy barrier, making transport much easier~\cite{Zhang_05}.  This model is very useful, especially for quickly calculating the energy barrier, and illuminating experimental current dependence on molarity~\cite{Bonthuis_06}. However, as we will review here, microscopic effects play a very important role
at these length scales. A detailed atomic picture of nanoscale features is thus essential to further understanding.

This paper is organized as follows.  In section~(\ref{sec:QP}) we discuss the interaction of ions and water leading to ionic quasi-particles, wherein ions and water combine to form the ionic atmosphere and hydration layers.  In section~(\ref{sec:QC}) we discuss the phenomenon of ionic "quantized" conductance, whereby the ionic current should experience jumps when the entrance radius of the pore is comparable to the average radius of the
hydration layers. In section~(\ref{sec:IB}) we consider the many-body effect of ionic blockade of nanopores of finite capacitance and discuss its relevance in both biological and synthetic pores. Finally, we conclude
in section~(\ref{sec:Conclusions}) with future possible directions to explore.

\section{Ionic Quasiparticles}
\label{sec:QP}

The charge of an ion surrounded by the rest of the water solution interacts readily with the highly polarizable water molecules and the other ions in the liquid. Two effects are particularly important to consider when dealing with ion transport: (1) the formation of an ionic atmosphere, and (2) the formation of hydration layers. Both can be considered as two ionic quasi-particle states: the bare ion charge is screened by ions
of different polarity (ionic atmosphere), or simply by the surrounding water molecules (hydration layers). As we will discuss below, these two quasi-particle states have different features and while the second always occur, the first may be neglected in many experiments that are now considered for DNA sequencing and detection. Nonetheless, for completeness, and because simple ohmic calculations of the ionic conductance in solution disregarding this   structure disagree with experimental values significantly~\cite{Kuyucak_94}, we do introduce it in the following.

\subsection{Ionic Atmosphere}
Ions in solution interacting with other ions cause the solution to be inhomogeneous, namely the densities of each ionic species are not uniform across the whole liquid.  A model to determine this inhomogeneity analytically was proposed by Debye and H${\rm \ddot{u}}$ckel~\cite{Debye_23}.  We give here a short account of its assumptions and main results, and refer the reader to  Ref.~\cite{Wright_07} for an extensive discussion.

The Debye-H${\rm \ddot{u}}$ckel model treats inhomogeneity due only to electromagnetic interactions between ions and further assumes that {\it i)} the electrolytes are completely dissolved into ions, {\it ii)} the ions do not interact with the liquid solution, and {\it iii)} the ions are assumed to be spherically symmetric and non-polarizable.

\begin{figure}
\centerline{
\includegraphics[width=0.70\linewidth]{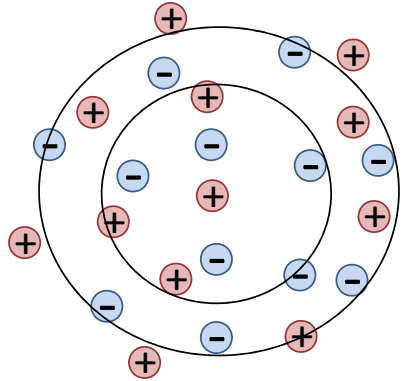}
}
\setlength{\columnwidth}{3.2in}
\caption{A cartoon picture of an ionic atmosphere.  In each successive radius around the $j$th ion there is more opposite charge.
\label{fig:IA} }
\end{figure}

Within these approximations each ion has an ionic atmosphere: a cloud of charge composed of other ions in the solution (see schematic in Fig.~\ref{fig:IA}). Due to electrostatic considerations (and under the above assumptions) it is more likely to find oppositely charged ions closer to a given ion - say the $j$-th ion - and similarly charged ions further from it. This difference in charge density is the source of the solution's inhomogeneity.
%
%
%

The key quantity to compute is the potential a distance $r$ away from the $j$-th ion due to the ionic atmosphere, call it $\phi_j^{IA}(r)$. This potential is then used to calculate the free energy difference caused by the inhomogeneity of the system. The total potential is
\begin{equation}
  \phi_j(r) = \phi^{ion}_j(r) + \phi^{IA}_j(r),
 \label{eq:pots}
\end{equation}
where $\phi^{ion}_j(r)$ is the potential due to the $j$-th ion, or $z_j e/4\pi\epsilon_0\epsilon_r r$, where $z_j$ is the integer charge of the ion, $\epsilon_0$ is the permittivity of free space, and $\epsilon_r$ the relative permittivity of the medium.

To find $\phi_j(r)$ one can use Poisson's equation
\begin{equation}
  \nabla^2\phi_j (r) = -\frac{1}{\epsilon_0\epsilon_r}\rho_j(r),
 \label{eq:ion_j}
\end{equation}
where $\rho_j(r)$ is the charge density, which is assumed to be spherically symmetric.

In the assumption of local equilibrium, a Maxwell-Boltzmann distribution provides the number of ions of species $i$ per unit volume as a function of distance $r$ from the $j$-th ion
\begin{equation}
  n_i' = n_i\exp\left(-\frac{z_ie\phi_j(r)}{k_BT}\right),
 \label{eq:Q_dens}
\end{equation}
where $n_i'$ is the concentration of ions of type $i$ at a given position, $n_i$ is the bulk concentration of ions of type $i$, $k_B$ is the Boltzmann constant, and $T$ is the temperature.

Since the sum of this term for each species is the the charge density $\rho_j(r)$, one obtains
\begin{equation}
  \nabla^2\phi_j (r) = -\frac{1}{\epsilon_0\epsilon_r}\sum_i \left[ n_i z_i e\exp\left(-\frac{z_ie\phi_j(r)}{kT}\right)\right],
 \label{eq:ion_j_gen}
\end{equation}
which generally requires a numerical solution. However, for symmetrical electrolytes expanding the exponential in equation (\ref{eq:ion_j_gen}) leads to the second-order equation
\begin{equation}
  \nabla^2\phi_j (r) =\kappa^2 \phi_j
 \label{eq:ion_j_approx}
\end{equation}
where
\begin{equation}
  \kappa^2 = \frac{e^2}{\epsilon_0\epsilon_r k T}\sum_i n_i z_i^2.
 \label{eq:kappa}
\end{equation}

Solving equation~(\ref{eq:ion_j_approx}) gives
\begin{equation}
  \phi_j (r) = \frac{z_je}{4 \pi\epsilon_0\epsilon_r}\frac{e^{\kappa r_0}}{1+\kappa r_0}\frac{e^{-\kappa r}}{r},
 \label{eq:psij}
\end{equation}
where $r_0$ is the center-to-center distance of closest approach for the ions.  Along with equation~(\ref{eq:pots}) this gives
\begin{equation}
  \phi^{IA}_j(r) = \frac{z_je}{4 \pi\epsilon_0\epsilon_r r}\left(\frac{e^{\kappa r_0}}{1+\kappa r_0}e^{-\kappa r}-1\right),
 \label{eq:psijIA}
\end{equation}
for the potential due to the ionic atmosphere.  This is the fundamental electrostatic description of the ionic atmosphere.  From here the charge density may be found
\begin{equation}
  \rho_j(r) = \frac{z_je\kappa^2}{4 \pi}\frac{e^{\kappa r_0}}{1+\kappa r_0}\frac{e^{-\kappa r}}{r}.
 \label{eq:dens}
\end{equation}
The quantity $1/\kappa$ is often defined as the effective radius of the ionic atmosphere, a length scale describing the ionic atmosphere's approximate size as measured from the $j$th ion.

The above analysis assumes that no external fields are present. However, the ionic atmosphere is a dynamic quantity, in the sense that the ``dressing'' of the bare ion charge has a finite lifetime. It is then affected by electric fields. The typical lifetime for an ionic atmosphere is 10$^{-8}$ s. Therefore, it generally does not survive in electric fields greater than 10$^4$ V/m. Therefore, for most of the experiments presently pursued in the field of DNA sequencing and
 detection this quasi-particle state bares little importance.

\subsection{Hydration Layers}
%
%
%
\begin{figure}
\centerline{
\includegraphics[width=0.8\linewidth]{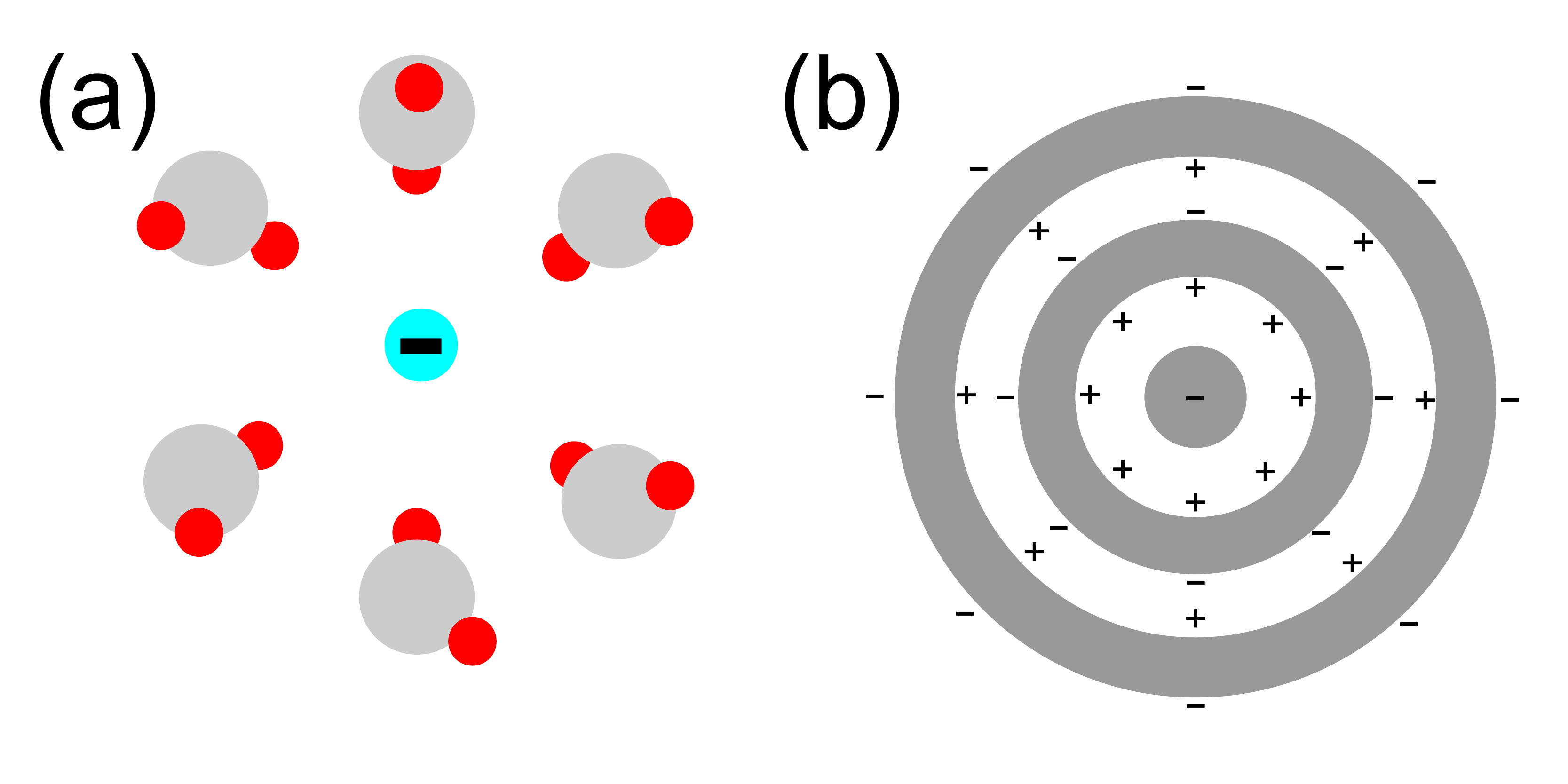}
}
\setlength{\columnwidth}{3.2in}
\caption{(a) Depiction of the first hydration layer around negative ion.  One hydrogen atom in each water molecule is oriented toward the negative ion.  (b) These hydration layers approximate charged concentric Gauss surfaces.  One may model them as such.
\label{fig:cartoon_CB} }
\end{figure}
An ion in an aqueous solution interacts strongly with the water molecules around it, forming hydration layers, or oscillations in water density due to alignment of water molecules with the surrounding electric field caused by the ion (see schematic in Fig.~\ref{fig:cartoon_CB}). These layers are discussed extensively in the literature~\cite{Hille_01, Marcus_94,Edsall_78,Ebbinghaus_07,Zhang_07}.  Early examination focused on determining the ionic radii, the lifetime of these bonds, and the interaction energies between ions and molecules~\cite{Hille_01,Otacki_93,Marcus_88}.  More recent studies have modeled the layers analytically~\cite{Marcus_94,Kuyucak_94}.  Comparisons with experimental results have been qualitatively - and in many instances also quantitatively - positive~\cite{Ebbinghaus_07,Zhang_07,Otacki_93,Marcus_88}. Recent advances in computing power make it possible to carry out all-atom molecular dynamics (MD)  simulations, and calculate the exact structure of these layers as well as their energetics.
%

%
%
%
\begin{figure}
\centerline{
\includegraphics[width=0.90\linewidth]{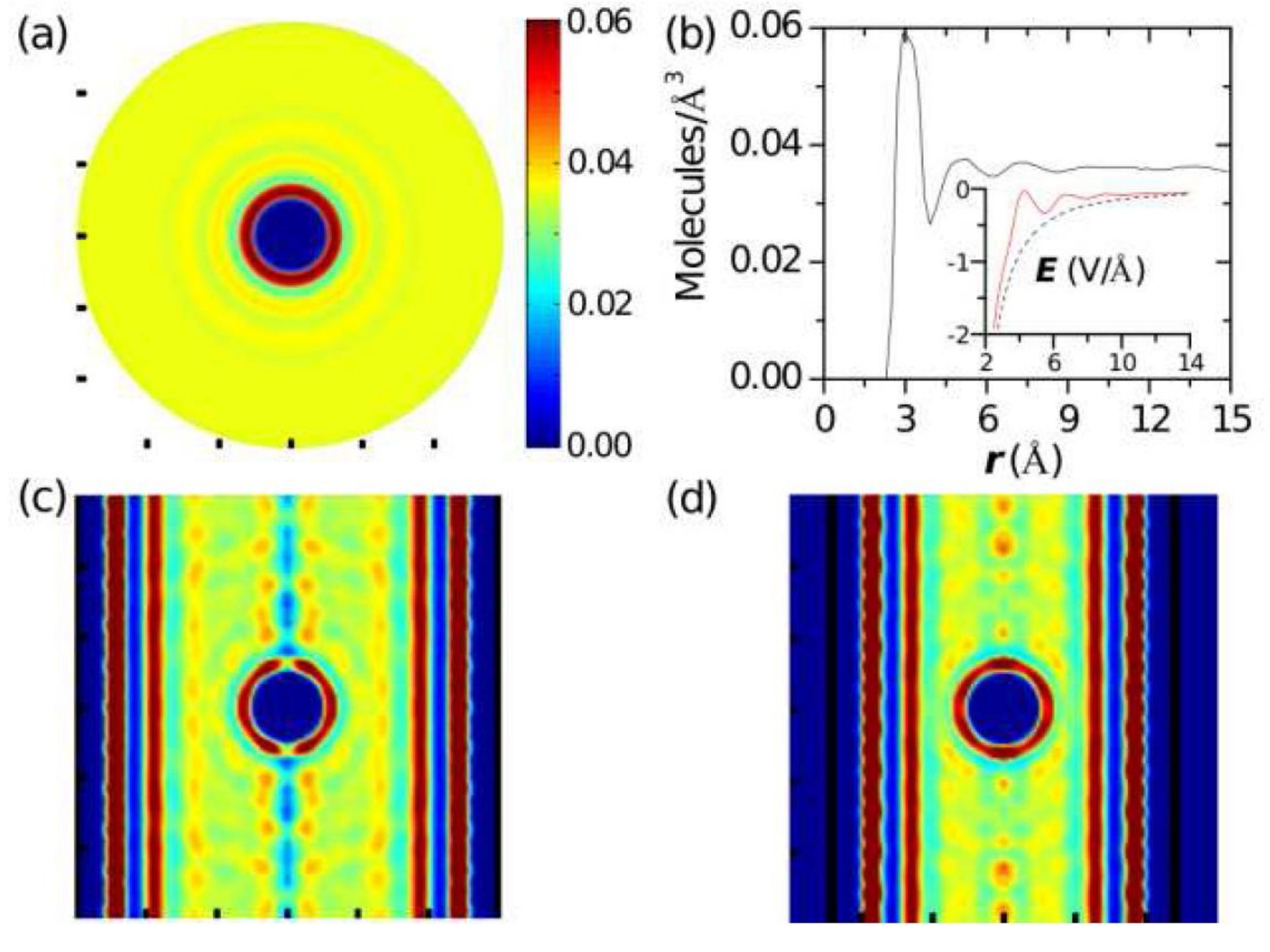}
}
\setlength{\columnwidth}{3.2in}
\caption{From Zwolak {\it et al.} \cite{Zwolak_09}:  (a) An illustration of water density around Cl$^-$ in bulk.
(b) Water density in relation to radial distance from Cl$^-$. The inset
shows the time-averaged radial electric field versus radial
distance from both the ion and water dipoles, shown as a red line, and from
just the ion, shown as a black dashed line. Water density surrounding Cl$^-$ inside
a (c) 15 \AA~radius pore and (d) 12 \AA~radius pore. The vertical thick black lines depict pore radius.  Reprinted with permission from Phys.\ Rev.\ Lett.\ {\bf 103}, 128102. Copyright 2009 by the American Physical Society.
\label{fig:MD_quasi_particle} }
\end{figure}
%

%
%
%
\begin{figure*}
\centerline{
\includegraphics[width=0.90\linewidth]{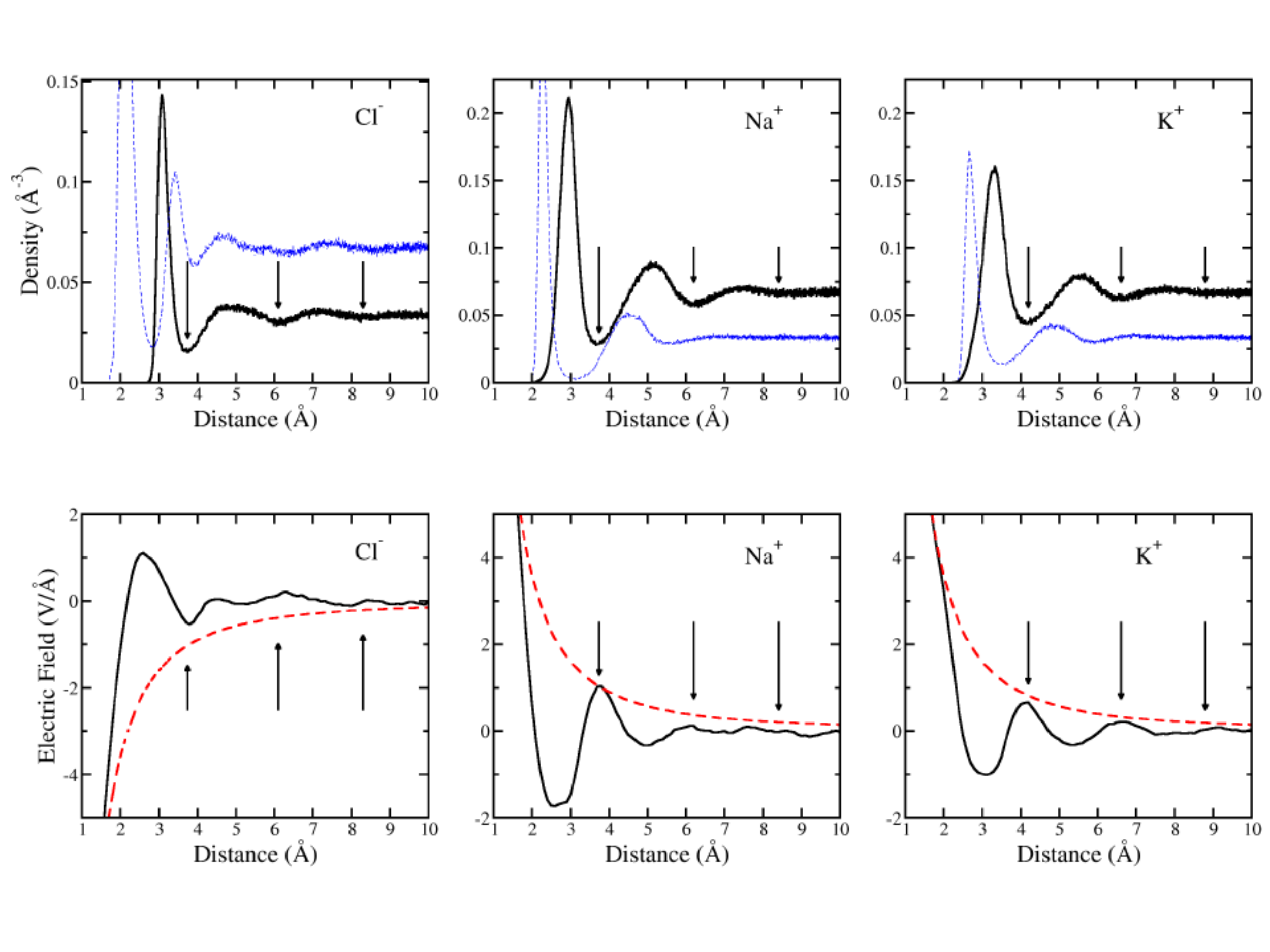}
}
\setlength{\columnwidth}{3.2in}
\caption{From Zwolak {\it et al.} \cite{Zwolak_10}: The upper panels depict water density oscillations versus distance for Cl$^-$, Na$^+$, and K$^+$ in bulk water. Solid black lines depict
the density calculated from the oxygen atom positions for Cl$^-$ and hydrogen atom positions for the anions. Minima in the density oscillations are depicted with arrows. Dashed blue lines depict the density calculated from the hydrogen
atom positions for Cl$^-$ and oxygen atom positions for the anions. In the lower panels dashed red lines depict the electric field due to the bare ion and solid black lines due to the ion plus partial charges on the water molecules, with the arrows indicating the
minima in the density oscillations.  Reprinted with permission from J.\ Phys.\ Condens.\ Matter\ {\bf 22}, 454126. © IOP Publishing. Reproduced by permission of IOP Publishing. All rights reserved.
\label{fig:MD_quasi_particle_more} }
\end{figure*}
Similar to quantum quasi-particles, these ionic quasi-particles acquire an effective mass and radius due to the ``dressing'' of water molecules around the bare ion charge.  Previous work has shown that these effective features are  dependent on temperature, with the number of associated water molecules in the quasi-particle decreasing with increasing temperature, thus lowering the effective mass \cite{Kuyucak_94}.  This leads to a temperature-dependent current since drift velocity is mass dependent.

Figure~(\ref{fig:MD_quasi_particle}) presents an example of recent MD simulation results for the density as a function of the distance away from the ion~\cite{Zwolak_09,Zwolak_10}.  Simulations were performed on various ions with few qualitative differences among various ionic species.  For Cl$^-$ shown in Fig.~(\ref{fig:MD_quasi_particle}) the water molecules orient themselves such that a hydrogen atom in the H$_2$O points toward the ion (see schematic in Fig~(\ref{fig:cartoon_CB}a)).  The molecules are pulled closer to the ion and are packed more tightly then in bulk water. The first layer has a peak density at approximately 3 \AA~for Cl$^-$, with each further layer at intervals of approximately 2 - 2.3 \AA.  These density peaks correspond to the location of a hydrogen or oxygen atom, depending on the sign of ion.  This is consistent with neutron diffraction and X-ray absorption measurements on Cl$^-$ solution~\cite{Otacki_93}.  The simulations show that the density returns to the value for bulk water (0.033 molecules/\AA) at around 10 \AA, which generally consists of three layers~\cite{Wright_07}.

Microscopic simulations also allow the calculation of the electric field generated by these quasi-particles. This field shows a similar oscillating pattern, see figure~(\ref{fig:MD_quasi_particle}), thus showing once again
that a continuum model would be totally insufficient to describe this type of features. Most interestingly, this field pattern can be approximated by a series of concentrically charged spheres, similar to the layers of an onion. This allows the calculation of the energy associated with each shell, see fig~(\ref{fig:cartoon_CB}b).

In calculating $U_{i}^o$, the energy within each hydration layer $i$, a Born solvation calculation can be used which  treats the hydration layers as concentrically charged spheres surrounding the ion.  Here the energy of the first layer is the difference in solvating the ion and solvating the ion and the first hydration layer, or generally for each layer $i$,
\begin{equation}
  U_{i}^o = {{e^2}\over{8\pi \epsilon_0}}\left({1\over\epsilon_{\rm p}}-{1\over\epsilon_{\rm w}}\right)\left({1\over R_{i}^{\rm O}}-{1\over R_{i}^{\rm I}}\right).
 \label{eq:U_i}
\end{equation}

In section~(\ref{sec:Predictions}) we will discuss how the hydration layers are predicted to affect ion transport in a nanopore system.

\section{Predictions}
\label{sec:Predictions}

\subsection{`Quantized' Conductance}
\label{sec:QC}
In the late 1980s a pressing experimental question concerned quantum ballistic transport: impurity-free transport through a quantum point contact, or a contact whose size approaches the Fermi wavelength of the electrons.  There was much debate as to how the conductance would behave as the contact width grew smaller.  Theorists considered whether there would even be resistance below a specific width~\cite{Landauer_57,Buttiker_85,Thouless_77,Imry_86}.  The discovery of quantized conductance, or the appearance of steps in the conductance as a function of the channel width, was a significant development~\cite{van_Wees_88}.  Here the conductance was found to be
\begin{equation}
  G = N \frac{2e^2}{h},
 \label{eq:QPC1}
\end{equation}
where $e$ is the electron charge, $h$ is the Planck constant, and $N$ is the number of modes allowed.  $N$ depends on the size of the point contact
\begin{equation}
  N \approx \frac{2W}{\lambda_F},
 \label{eq:QPC2}
\end{equation}
where $W$ is the width of the point contact, and  $\lambda_F$ is the Fermi wavelength~\cite{Di Ventra_08}.

Since the `size' of the electron is on the same order as the contact, the conductance cannot take on any size but instead is quantized.  This result has many analogies, for example the transmission of monochromatic light through thin slits and the thermal conductance of a quantum point contact~\cite{van_Houten_96}.  Similarly, an ionic quasiparticle transversing a channel roughly the same size as its effective radius is predicted to acquire a `quantized' conductance~\cite{Zwolak_09,Zwolak_10}.

When put into a nanopore environment, the behavior of the hydration layers is extremely dependent on the radius of the pore.  For large pores, the layers are relatively unperturbed.  When the size of the pore approaches the size of the well-defined hydration layers they will partially break, as they cannot physically fit in the pore.  As the ion approaches the pore the excess hydration layers are shed.  This shedding will inflict an energy penalty on the translocation, resulting in ionic quantized conductance.  MD simulations by Zwolak {\it et al.}, see figure~(\ref{fig:MD_quasi_particle})~(c) and (d), show interference patterns from the interactions of the pore wall and the outermost hydration layers of a ionic quasiparticle immersed in a pore.  A model was devised to explain these effects analytically.  In this model only parts of the hydration layer allowed in the pore remain.  Here the internal energies of the hydration layers are estimated as
\begin{equation}
  U_{i} = f_i U_{i}^o,
 \label{eq:U_i_sing_hydr}
\end{equation}
where $f_i$ is the fraction of hydration layer remaining after breakage and $U_{i}^o$ is the internal energy difference between the hydration layers as parts of a quasiparticle and the water in bulk.  Therefore the energy barrier is due to the energy required to peel off part of the hydration layer.  One can calculate the area remaining of hydration layer $i$ when an ion translocates through a pore as

\begin{equation}
  S_{i}^o = 2\Theta \left(R_{i}-R_{\rm p}\right)\int_0^{2\pi}{d\phi}\int_0^{\theta_{c}}{d\theta R_{i}^2\sin{\theta}},
 \label{eq:S}
\end{equation}
where $R_i$ is the radius of the pore and
\begin{equation}
  \theta_{c}=\sin^{-1}{\left(R_{\rm p}/R_{i}\right)}.
 \label{eq:thetacu}
\end{equation}

This is easily solved as
\begin{equation}
  S_{i}^o = 4\pi R_i^2\left(1 - \sqrt{1-\left(\frac{R_{\rm p}}{R_{i}}\right)^2}\right),
 \label{eq:S}
\end{equation}
leading to the fraction of surface area remaining
\begin{equation}
  f_{i}\left(R_{\rm p}\right) = 1 - \sqrt{1-\left(\frac{R_{\rm p}}{R_{i}}\right)^2},
 \label{eq:fraction}
\end{equation}
which leads to an internal energy difference of
\begin{equation}
  \Delta U\left(R_{\rm p}\right) = \sum_{i}\left(f_{i}\left(R_{\rm p}\right) - 1\right)U_{i}^o
 \label{eq:deltaU}
\end{equation}
for each hydration layer $i$.  In order to calculate the free energy, an entropic contribution must be included.  By taking one ion out of the solution and inserting it into the pore there is a change in entropy on the order of
\begin{equation}
  \Delta S = k_B \ln{\left(V_{\rm p}n\right)},
 \label{eq:deltaS}
\end{equation}
where $V_{\rm p}$ is the volume of the pore and $n$ is the bulk ionic concentration.  This gives a change in free energy
\begin{equation}
  \Delta F = \Delta U - T \Delta S,
 \label{eq:deltaF}
\end{equation}
or
\begin{equation}
  \Delta F = \sum_{i}\left(f_{i}\left(R_{\rm p}\right) - 1\right)U_{i}^o - T \Delta k_B \ln{\left(V_{\rm p}n\right)}.
 \label{eq:deltaF}
\end{equation}
Step-like features are then produced in the free energy as a function of pore radius, similar to those in quantum point contact experiments~\cite{Zwolak_09,Zwolak_10}.  These are the features that lead to quantized conductance through the pore.

\begin{figure}
\centerline{
\includegraphics[width=0.80\linewidth]{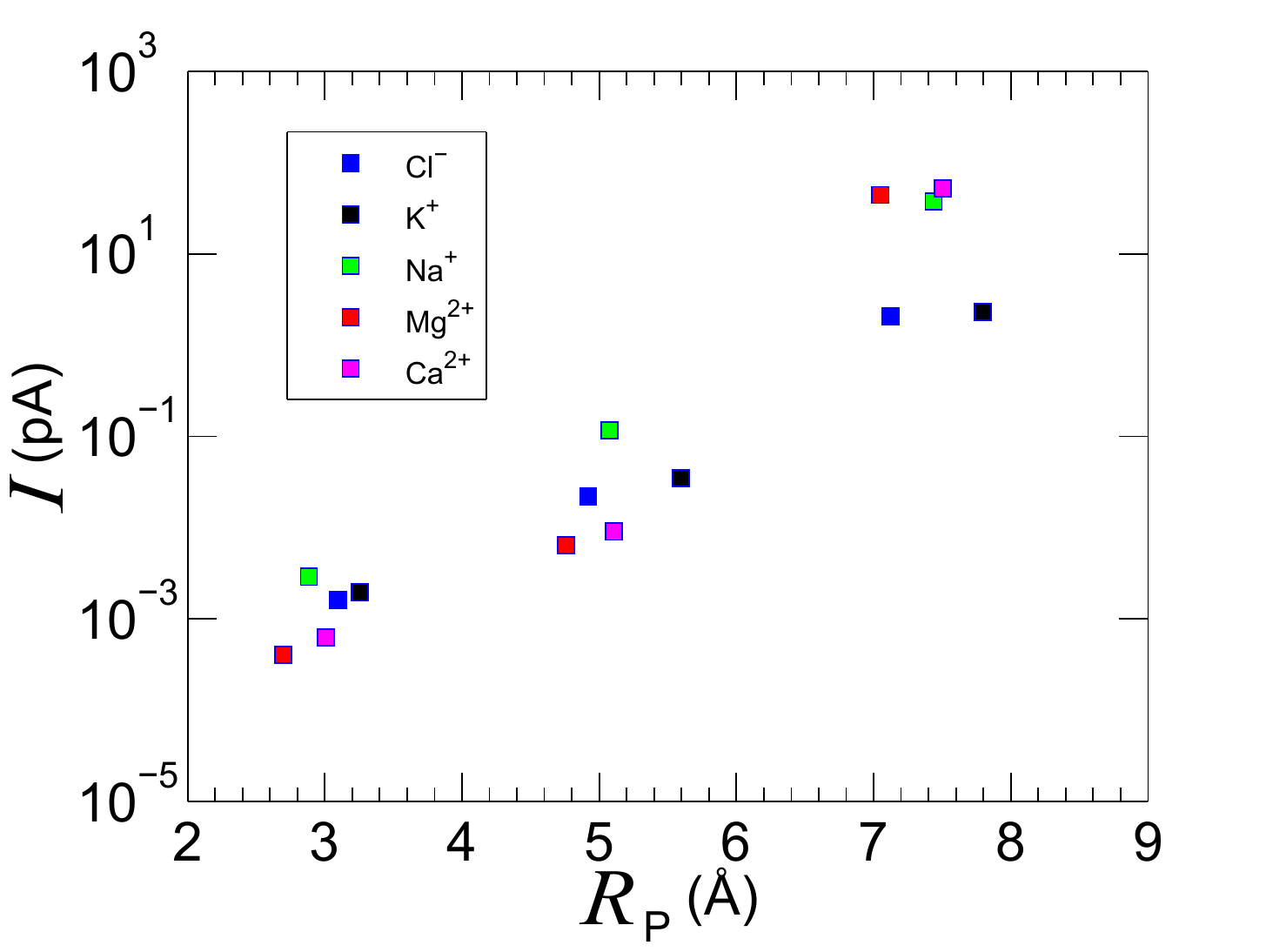}
}
\setlength{\columnwidth}{3.2in}
\caption{Current vs. effective pore radius for a field of
1mV/\AA.  Values show locations where steps in current occur.  Here the effective pore radius approximately equals the radius of the hydration layers.
\label{fig:ICB} }
\end{figure}

To calculate the current through the pore based on these energy barrier calculations one can employ the 1-d Nernst-Planck equation,
\begin{equation}
  J_\nu = -q_\nu D_\nu \left[\frac{{\rm d}n_\nu \left(z\right)}{{\rm d}z} + \frac{q_\nu}{k_BT}n_\nu\left(z\right)\frac{{\rm d}\Phi_\nu \left(z\right)}{{\rm d}z}\right],
 \label{eq:NP}
\end{equation}
for the steady state.  Here $J_\nu$ is the current density, $z$ is the axial coordinate, $l$ is the length of the pore, $D_\nu$ is the diffusion coefficient and $\Phi_\nu(z)$ is the position-dependent potential (including all interactions that affect energy in the pore).  Here, it is assumed that the density on either side of the pore remains constant and equal.  Since the pore is of high conductance this is an appropriate approximation.  So multiplying by $e^{q_\nu \phi_\nu\left(z\right)/k_B T}$,
\begin{equation}
  J_\nu e^{q_\nu \phi_\nu\left(z\right)/k_B T} = -q_\nu D_\nu \frac{{\rm d}}{{\rm d}z}\left[n_\nu\left(z\right)e^{q_\nu \phi_\nu\left(z\right)/k_B T}\right],
 \label{eq:Jvint}
\end{equation}
and integrating leads to the current density
\begin{equation}
  J_\nu  = -q_\nu D_\nu {n_R e^{q_\nu \phi_\nu\left(z\right)/k_B T}-n_L e^{q_\nu \phi_\nu\left(z\right)/k_B T}\over\int_0^l {\rm d}z e^{q_\nu \phi_\nu\left(z\right)/k_B T}}.
 \label{eq:Jv}
\end{equation}
Making some more simplifications: {\it i)} the electrostatic potential drops linearly over the pore and {\it ii)} the potential barrier of other contributions is essentially constant over the pore.  This leads to a potential of
\begin{equation}
  \Phi_\nu(z) = z {V\over l} + {\Delta F_\nu\over q_\nu}.
 \label{eq:potentialNP}
\end{equation}
Integration leads to
\begin{equation}
  J_\nu  = -\frac{q_\nu^2 n_0 D_\nu V}{lk_BT}e^{-\Delta F_\nu/k_B T}.
 \label{eq:Jv}
\end{equation}
Finally, calculating the current from current density, one has
\begin{equation}
  I_\nu  = -2\pi R_{\rm p}^2J_\nu = I_{\nu 0}e^{-\Delta F_\nu/k_B T}.
 \label{eq:Jv}
\end{equation}
Determination of the current in Refs~\cite{Zwolak_09,Zwolak_10} was made in this manner using results of the free energy.  Figure (\ref{fig:ICB}) gives a plot of current vs. pore radius ($R_{\rm P}$).  Energetic barriers lead to steep drops in the current when $R_{\rm P}$ equals the radii of the hydration layers, the locations of which are as expected (section~(\ref{sec:QP})).

%
%
%

The potential impact of noise is important to any experimental realization.  Noise could affect the visibility of these current drops.  This has been addressed as well~\cite{Zwolak_09, Zwolak_10}.  The two main sources of noise that were investigated are: {\it i)} Gaussian fluctuations in the free energy barrier and {\it ii)} Gaussian fluctuations in the effective pore radius. The first one would be due to the dynamic nature of the hydration layers and the second one to possible charges at the pore walls and fluctuations in water molecule density in the pore.

The relative current noise due to fluctuations in the free energy barriers was found to be~\cite{Zwolak_09, Zwolak_10}
\begin{equation}
  \Delta I_{rel} \approx \frac{\sigma}{k_B T},
 \label{eq:Noise_EB}
\end{equation}
where $\sigma$ is the standard deviation of the noise.  Therefore this current noise increases with the broadness of the noise distribution.  These fluctuations will decrease the effective energy barrier and increase the current, making the drops smaller but not eliminating them completely.  However, at higher $\sigma$ the current increases and the barrier grows smaller with the third barrier being the largest, which is not very likely.

Alternatively, fluctuations in the effective size of the pore will also result in a smoothing out of the peaks.  However, these fluctuations also lead to an interesting peak in the relative current noise:
\begin{equation}
  \Delta I_{rel}^{peak} \approx \frac{1}{2}e^{\Delta F_h / 2 k_B T},
 \label{eq:Noise_EPS}
\end{equation}
where $\Delta F_h$ is the energy barrier for a hydration layer.  This leads to an exponential peak in relative noise that exists in environments with large amounts of noise and will occur when the effective pore radius is close to the size of hydration layer.  Qualitatively this occurs when the hydration layer radius is close to the pore radius, therefore fluctuations have the effect of `opening' or `closing' the pore leading to very different currents.

\begin{figure}
\centerline{
\includegraphics[width=0.8\linewidth]{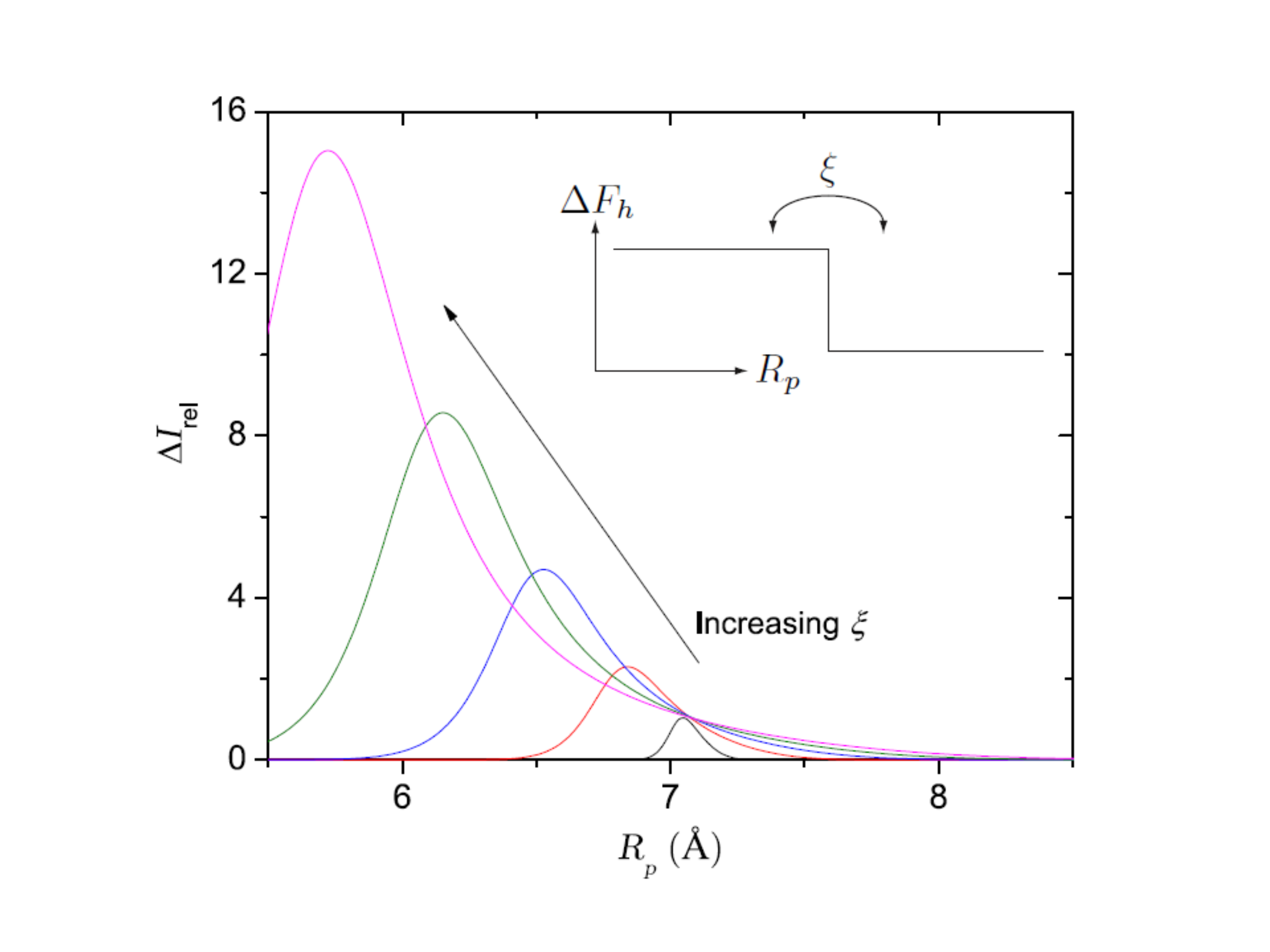}
}
\setlength{\columnwidth}{3.5in}
\caption{From Zwolak {\it et al.}\cite{Zwolak_10}: Relative current noise caused by structural fluctuations
in the effective pore radius. The inset shows the approximate change in free energy,
$\Delta F_h$, as a function of effective pore radius in proximity
to a hydration layer structure. Fluctuations between the high and low energy
states are caused by noise in the pore radius. Here the third hydration layer radius of Cl$^-$ is taken,
$R_h$ = 7.1 $\AA$. $\Delta F_h$ is explained in the text. Fluctuation
strength from right to left is $\xi$ = 0.05; 0.15; 0.25; 0.35; 0.45 $\AA$.  Reprinted with permission from J.\ Phys.\ Condens.\ Matter\ {\bf 22}, 454126. © IOP Publishing. Reproduced by permission of IOP Publishing. All rights reserved.
\label{fig:noise_peak} }
\end{figure}

Figure~(\ref{fig:noise_peak}) shows the relationship between the relative current noise and effective pore radius for different values of $\xi$, the standard deviation of the fluctuations in pore size.  The effective pore radius where this noise peaks decreases with $\xi$.  This presents an intriguing way to test for quantized conductance that is useful in many different environments.

%
%
%

\begin{figure}
\centerline{
\includegraphics[width=0.70\linewidth]{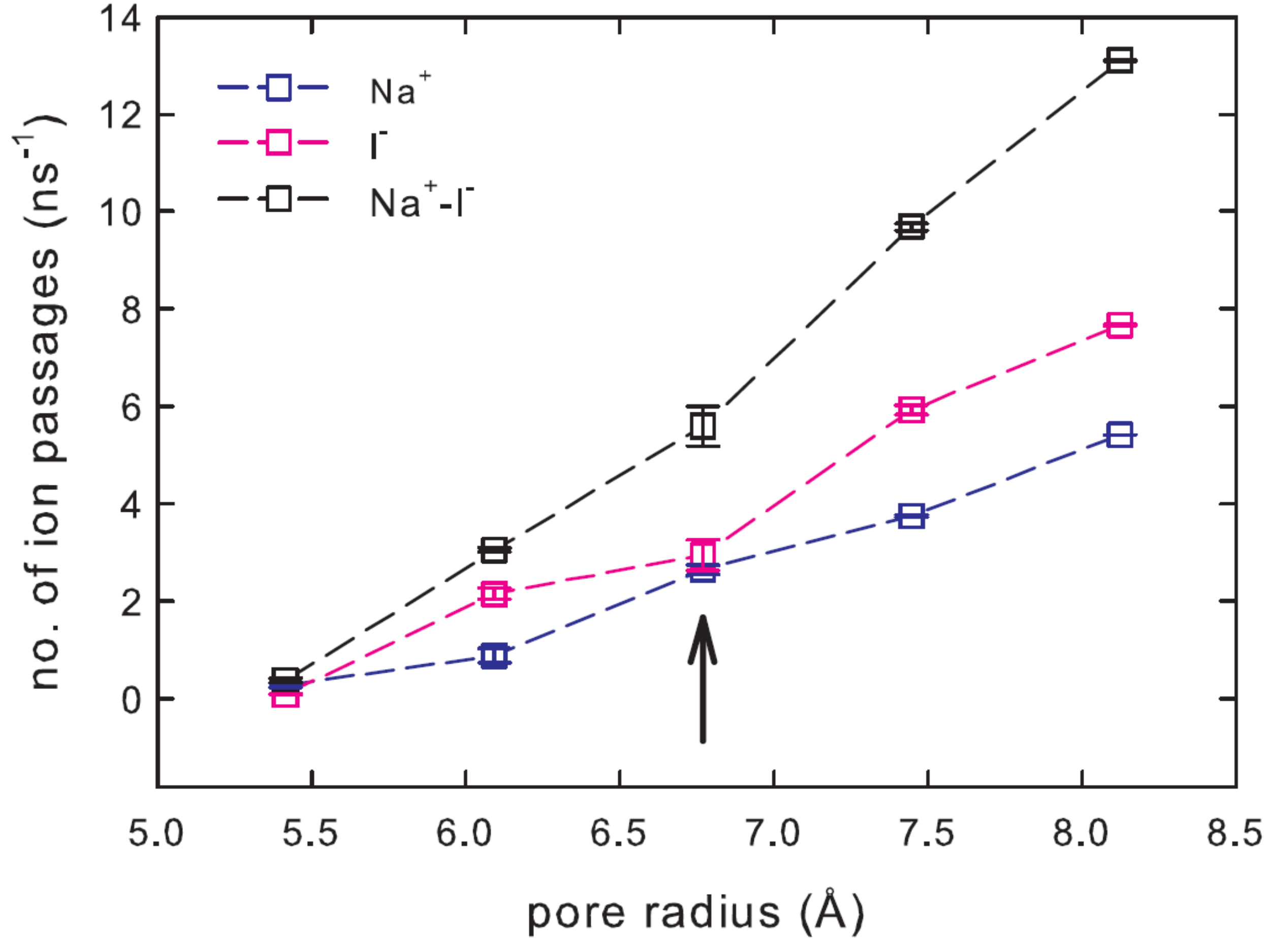}
}
\setlength{\columnwidth}{3.2in}
\caption{From Beu et al.~\cite{Beu_11b}:  Number of ion passages per nano second as a function of pore radius for NaI solution subject to an axially-symmetric external electric field as a function of pore radius in CNT.  Model is for 30 ions and field strength of 0.02 V/\AA.  Non-linear behavior is noted by arrow.  (Note: when corrected for half of graphite inter-plane distance $\sigma_{\rm graph}$ = 1.674 \AA~non-linear behavior occurs at $R_{\rm p}$ = 5.10 \AA.)  Reprinted with permission from J.\ Chem.\ Phys.\ {\bf 135}, 044516. Copyright 2011, American Institute of Physics.
\label{fig:beu} }
\end{figure}

Since the reporting of these results, possible quantized conductance patterns have been also observed in all-atom MD simulations of similar environments~\cite{Beu_11b,Zhao_10}.  For instance, Beu has observed a current discontinuity involving carbon nanotubes (CNT)~\cite{Beu_11b}.  In figure~(\ref{fig:beu}), the current vs. pore radius of his simulations is reported, showing a discontinuity at $R_{\rm p}$ = 5.10 \AA~in Na$^+$ as expected for the breaking of the the second hydration layer, see figure~(\ref{fig:ICB}).  Additionally Zhao {\it et al.} used MD simulations for CNT systems with various radii and found transport results consistent with energetic barriers for the breaking of hydration layers~\cite{Zhao_10}.  However, experimental confirmation of this phenomenon is still needed.
%
%
%
%

\subsection{Coulomb Blockade}
\label{sec:IB}
\begin{figure}
\centerline{
\includegraphics[width=0.70\linewidth]{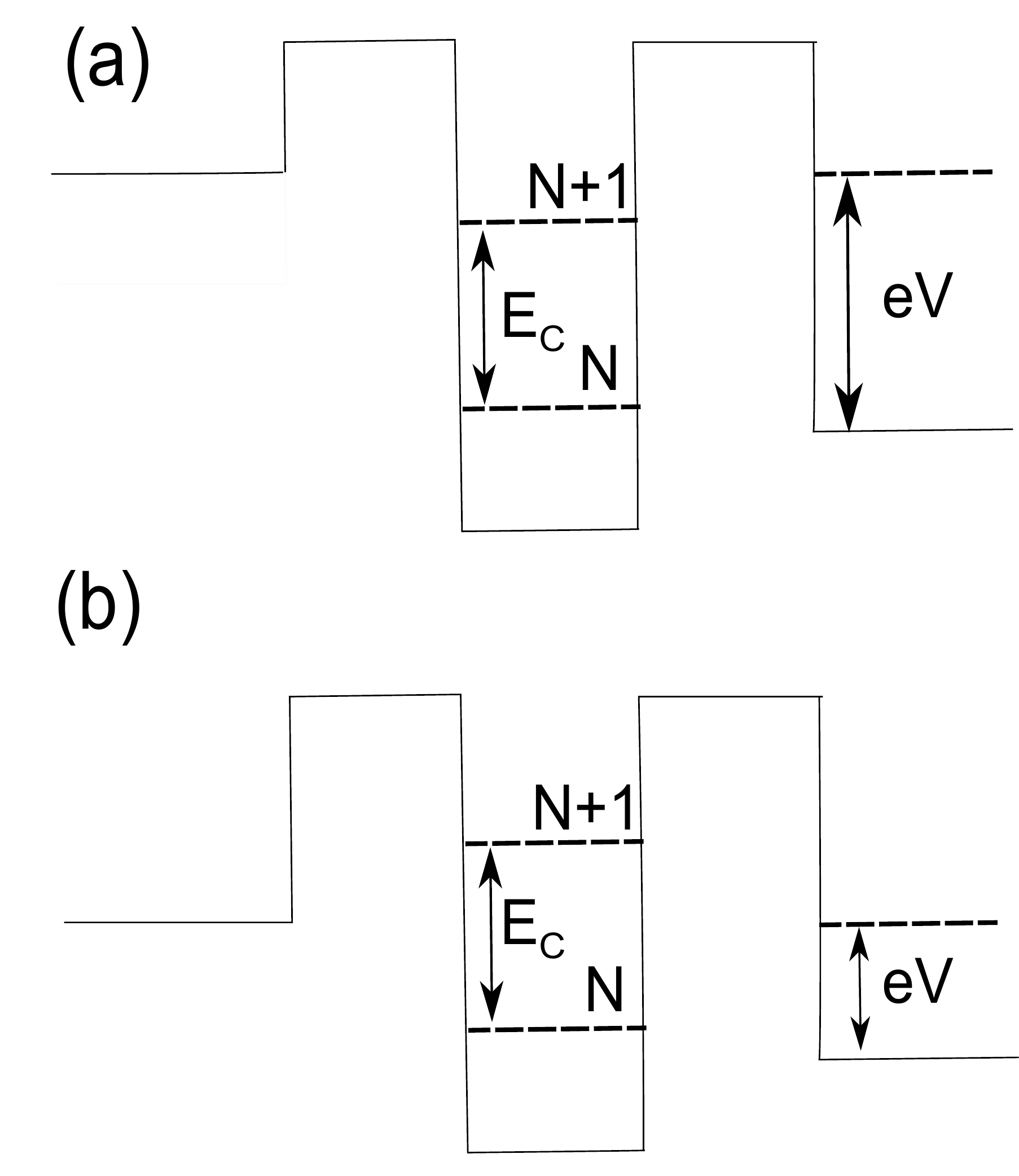}
}
\setlength{\columnwidth}{3.2in}
\caption{Energy diagram considering a central region weakly connected to electrodes through tunneling junctions. Dotted lines represent energy of states of $N$ and $N+1$ electrons.  (a) Bias is large enough to overcome the capacitative barrier and transport is allowed.  (b) Bias is not enough to overcome the capacitative barrier leading to coulomb blockade.
\label{fig:CB}}
\end{figure}

Coulomb blockade is a phenomenon where tunneling across a junction is either enhanced or suppressed for certain energies as a result of the quantization of charge and Coulomb effects~~\cite{Averin_86}.  Consider the energy diagram in figure~(\ref{fig:CB}), where a central region is connected to two electrodes via tunneling junctions.  The central region can be assumed to have a capacitance $C$.  If this region contains a charge $Q$ there is an electrostatic energy of $E = Q^2/2C$.  With the addition of an extra electron this energy will change to
\begin{equation}
  E = \frac{\left(|Q|-|e|\right)^2}{2C}.
 \label{eq:CB1}
\end{equation}
Energetically this new state is favorable when the new energy is less than or equal to the existing value or
\begin{equation}
   \frac{\left(|Q|-|e|\right)^2}{2C} \leq \frac{Q^2}{2C}.
 \label{eq:CB2}
\end{equation}
Reducing this and inserting the relationship for capacitance $C = Q/V$,
\begin{equation}
   |V| \ge \frac{|e|}{2C}.
 \label{eq:CB3}
\end{equation}
Therefore tunneling and thus transport are likely only when the bias exceeds the charging energy of the central region as outlined in figure~(\ref{fig:CB}).  Otherwise there will be virtually no electron transport or Coulomb blockade.  This condition continues for multiple electrons, thus creating steps in bias needed to overcome each new blockade.  For further discussion please see reference texts~\cite{Di Ventra_08}.

%
%
%
\begin{figure}
\centerline{
\includegraphics[width=0.70\linewidth]{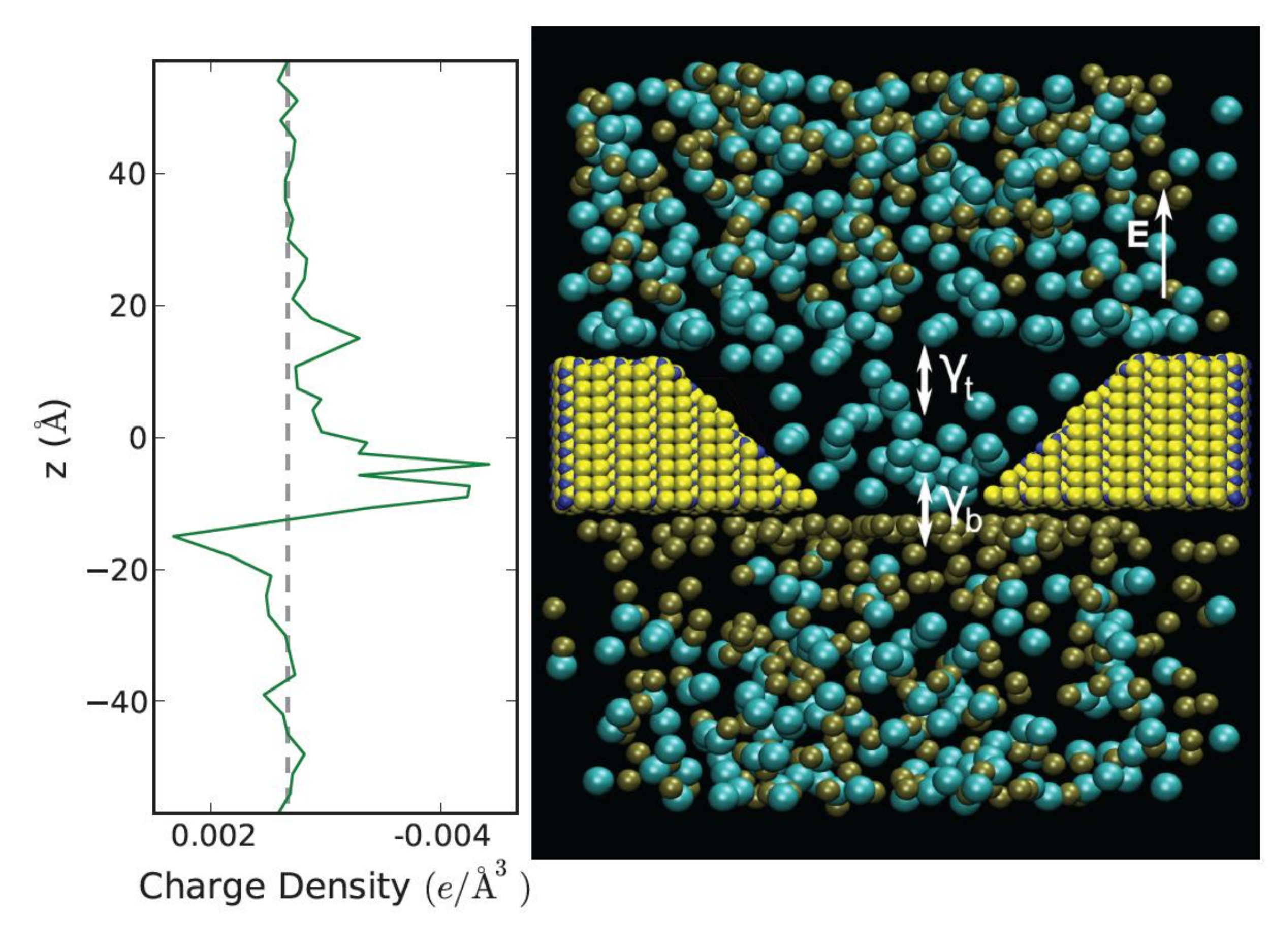}
}
\setlength{\columnwidth}{3.2in}
\caption{From Krems and Di Ventra \cite{Krems_13}: The snapshot on the right illustrates a molecular
dynamics simulation of 1M KCl solution subject to an
electric field and translocating through a cone-shaped Si$_3$N$_4$
pore (yellow and blue atoms) at a pore slope
angle of 45 with a bottom opening of 7\AA~a radius and thickness
of 25\AA. Cl$^-$ ions are shown in aqua and K$^+$ ions are shown in brown. The rates of ion
transfer at the bottom and top openings are indicated by $\gamma_t$ and
$\gamma_b$, respectively. This figure was generated using a field of 5.0 kcal=(mole)
allowing a simpler visualization of ion accumulation. The graph on the left shows
the net charge density corresponding to the right panel configuration.
The build-up of Cl$^-$ ions inside the pore is apparent, with the K$^+$ ions located mostly
outside the pore.  Reprinted with permission from J.\ Phys.\ Condens.\ Matter\ {\bf 25}, 065101. © IOP Publishing. Reproduced by permission of IOP Publishing. All rights reserved.
\label{fig:CB_simulation} }
\end{figure}

Similarly, a consequence of the formation of ionic quasi-particles is the {\it ionic Coulomb blockade} (ICB), caused by screened ion-ion interactions~\cite{Krems_13}.  This effect occurs as ions flowing through a nanopore via an external bias build up in the nanopore, hindering the flow of other ions.  A picture resulting from MD simulations is provided in figure~(\ref{fig:CB_simulation}).  This model is highly dependent on pore geometry.  For this discussion the top pore opening is assumed to be larger then the bottom.  Also, this effect is independent of charge polarity, however for discussion a negative charge is assumed.  Here, anions heading into the larger top opening get trapped when exiting the narrower bottom opening. Cations heading in the opposite direction accumulate outside the bottom, or smaller, opening.  These concentrated ions form a capacitor with energy barrier $Q^2/2C$, where $Q$ is the average number of ions in the channel and $C$ is the capacitance of this particular geometry.  An approaching ion will be blocked if its kinetic energy, a combination of thermal energy and drift kinetic energy, is less than that of this barrier.  This blockade is analogous to the Coulomb blockade demonstrated in quantum dot systems coupled to two electrodes~\cite{Amman_89,van Wees_91,Beenakker_91,Foxman_93,Kouwenhoven_01}.   In these systems the tunneling resistance between electrodes is much larger than the quantum of resistance.  In a similar way ICB is expected when the nanopore is weakly coupled to at least one of the ion reservoirs.  In other words, the time required for an ion to translocate
 through the pore is much longer than the time required to drift by in the absence of the pore.   This demands that the `contact' ionic resistance of the pore with at least one reservoir to be much larger than the resistance offered without the pore.  This will occur, as noted in the previous section, when the size of the pore approaches the `quantum' regime, or when it approaches the size of the hydration layer radii of an ionic quasiparticle.

While this analogy is strong, there is one major difference.  Ions are not limited by Fermi statistics as this is not a system in the quantum regime.  Therefore there is no maximum number of ions that can build up in the pore
other than that imposed by its capacitance.  The Coulomb blockade effect is then expected to have non-linear dependence on the ionic concentration, which in this analogy could be thought of as the control `gate voltage'.  There should, however, be ohmic dependence on the bias that drives the ions~\cite{Krems_13}.

There are previous reports of non-ohmic behavior in nanopore systems~\cite{Wright_07,Powell_07,Cruz-Chu_09,Coronado_80,Lu_94,Rostovtseva_98}. However, different physical circumstances explain most of these cases, primarily inter-ionic activity associated with the ionic atmosphere ~\cite{Powell_07,Cruz-Chu_09,Coronado_80}, which, as mentioned in section \ref{sec:QP}, is a dynamic phenomenon not applicable to this work. Furthermore, there are few examples of a decrease in current with increased concentration. This type of behavior was predicted to occur in systems due to ionic interactions with bulk ions in the reservoir~\cite{Nadler_04}. Again this effect has a different root cause and does not predict such a dramatic decrease. Other decreases in conductance have been observed and warrant further study~\cite{Lu_94,Rostovtseva_98}.
%
%
%

A simple analytic model, mirrored from its quantum equivalent, has been proposed to capture the main qualitative aspects of this phenomenon~\cite{Krems_13}.  As with quantum dot Coulomb blockade, this model uses a rate equation.  Referring to figure~\ref{fig:CB_simulation} the model defines two rates for ion transfer: (1) $\gamma_t$, the rate at the top opening in the figure (wide opening), and (2) $\gamma_b$, the rate at the bottom opening in the figure (neck opening).  The rates are related as such:
\begin{equation}
            \gamma_b =\alpha \gamma_t ,\quad  0 < \alpha < 1.
\label{eq:gamb}
\end{equation}
Since the top opening is less resistant to flow due to a lower level of shedding of hydration layers, transfer into the top opening is greater than that of the bottom opening.  The parameter $\alpha$ takes this into account.

Here $\gamma_t$ is defined in the usual way:
\begin{equation}
            \gamma_t =A_t\mu n_0 E,
\label{eq:gamt}
\end{equation}
where $A_t$ is the top area of the pore, $\mu$ is the ionic mobility, $n_0$ is the ionic concentration, and $E$ is the electric field assumed to be constant over the pore $E=V/d$, where $V$ is the electric potential across the pore and $d$ is the pore length.

To have an analytical solution, only two `ionic states' are assumed: (1) a state with only the background charge in the pore with transition probability $P_0$ and (2) a state with this background charge plus one additional ion with transition probability $P_1$.  The background charge is assumed to be $\lfloor n_0 \Omega_{\rm p} \rfloor$, here $\Omega_{\rm p}$ is the volume of the pore and  $\lfloor ... \rfloor$ is the floor function.  This simple model encompasses the qualitative features of this theory, namely a capacitive energy barrier to transport, while remaining analytically soluble.  Additionally, it provides an excellent means of comparison with MD simulations.

The rate equation describing the change of the transition probabilities is
\begin{equation}
            \frac{dP_0}{dt} = \Gamma_{1\rightarrow 0}P_1 - \Gamma_{0\rightarrow 1}P_0,
\label{eq:grate}
\end{equation}
where $\Gamma_{1\rightarrow 0}$ is the transition rate of going from state 1 to
state 0, and $\Gamma_{0\rightarrow 1}$ is the reverse process.

The transition rate from state 1 to 0 is given by
\begin{equation}
             \Gamma_{1\rightarrow 0}= \gamma_t \exp\left(-\epsilon_C/k_B T\right),
\label{eq:Gamma1to2}
\end{equation}
where $\epsilon_C$ is a single-particle capacitive energy barrier, $k_B$
the Boltzmann constant, and $T$ the temperature.  As covered in the last section, the model uses the Nernst-Planck equation in the steady state, equation~(\ref{eq:NP}), to calculate the current, and thus the rate.  This transition rate calls for a strong energetic barrier to transport, in agreement with the qualitative picture.

Here the single-particle capacitive energy is expressed as
\begin{equation}
             \epsilon_C = e^2 \frac{2\lfloor n_0 \Omega_{\rm p} \rfloor + 1}{2C},
\label{eq:epsC}
\end{equation}
where $e$ is the elementary charge and $C$ is the capacitance
of the pore.

The exact capacitance will be system dependent, however it must depend linearly on the surface areas of both the top and bottom opening of the pore.  The natural parameter here is $\alpha$  as defined in equation~(\ref{eq:gamb}). The model then assumes a value
$C = \alpha C_0$ with $C_0$ a reasonable experimental value
for the capacitance.

The rate $\Gamma_{1\rightarrow 0}$ is
\begin{equation}
             \Gamma_{1\rightarrow 0}= \gamma_b.
\label{eq:Gamma1to0}
\end{equation}
There is no capacitive barrier for this process, so the rate required to take a charge out of the pore remains steady.

Fundamentally, there is also a finite rate of
ion movement in the direction opposite of that determined
by the electric field. This rate would contribute to both transition rates
$\Gamma_{0\rightarrow 1}$ and $\Gamma_{1\rightarrow 0}$. However, this process is exponentially
suppressed for the biases considered here, and can
be safely ignored.
%
%
\begin{figure}
\centerline{
\includegraphics[width=0.70\linewidth]{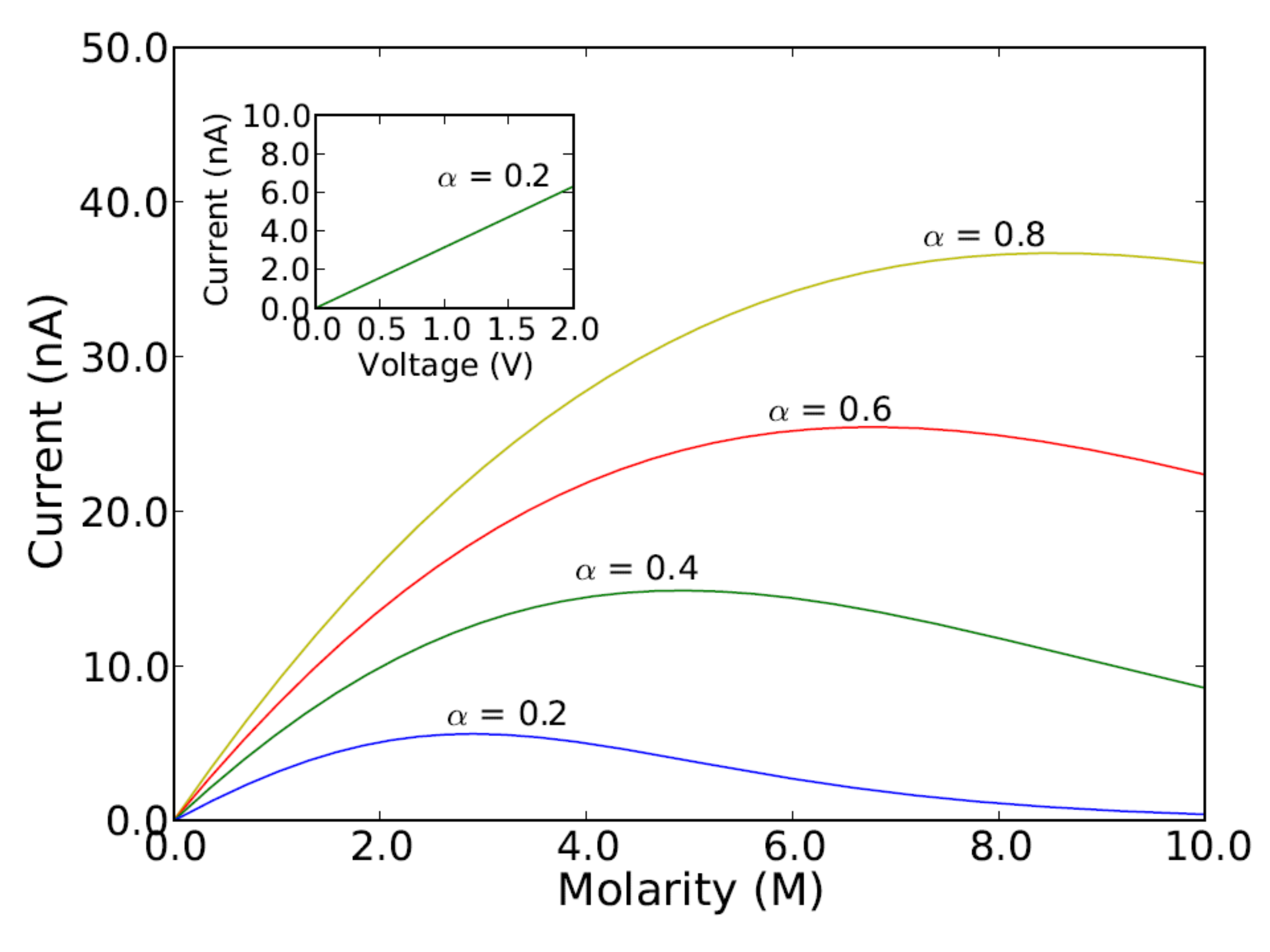}
}
\setlength{\columnwidth}{3.2in}
\caption{From Krems and Di Ventra \cite{Krems_13}: Current from equation~(\ref{eq:Ib2}) as a function
of the ion concentration at a fixed voltage of V = 1.0 V for
various pore neck openings as represented by the parameter $\alpha$ (larger $\alpha$ implies larger neck opening). The inset shows the
current, at a concentration of 1.0 M, as a function of bias for
 $\alpha$ = 0.2.  Reprinted with permission from J.\ Phys.\ Condens.\ Matter\ {\bf 25}, 065101. © IOP Publishing. Reproduced by permission of IOP Publishing. All rights reserved.
\label{fig:Ib} }
\end{figure}
Assuming steady state, $dP/dt = 0$, and using equation~(\ref{eq:grate}), and the
equivalent equation for $P_1$,  one can find the steady-state
probabilities~\cite{Krems_13}
\begin{equation}
             P_0 = \frac{\Gamma_{1\rightarrow 0}}{\Gamma_{0\rightarrow 1} + \Gamma_{1\rightarrow 0}} = \frac{\gamma_b}{\gamma_t \exp\left(-\epsilon_C/k_B T\right) + \gamma_b}
\label{eq:P0}
\end{equation}
and
\begin{equation}
             P_1 = \frac{\Gamma_{0\rightarrow 1}}{\Gamma_{0\rightarrow 1} + \Gamma_{1\rightarrow 0}} = \frac{\gamma_t \exp\left(-\epsilon_C/k_B T\right)}{\gamma_t \exp\left(-\epsilon_C/k_B T\right) + \gamma_b}.
\label{eq:P1}
\end{equation}
Finally, the current at steady state is the same everywhere
and is evaluated across the neck of the pore as
\begin{equation}
            I_b = e\left(\Gamma_{0\rightarrow 1}^bP_1-\Gamma_{1\rightarrow 0}^bP_0\right),
\label{eq:Ib1}
\end{equation}
where the superscript $b$ indicates that only the
terms corresponding to the bottom part of the transition
rates are retained. In the present case, this gives
\begin{equation}
            I_b = e\alpha\gamma_t\left(\frac{\exp\left(-\epsilon_C/k_B T\right)}{\exp\left(-\epsilon_C/k_B T\right)+\alpha}\right),
\label{eq:Ib2}
\end{equation}
with $\gamma_t$ and $\epsilon_t$ given by equations~(\ref{eq:gamt}) and ~(\ref{eq:epsC}), respectively.

This equation reflects this model's expectations.  For a fixed gate voltage, there is indeed a non-linear response with respect to molecular concentration.  As shown in figure~(\ref{fig:Ib}) initially  the current increases linearly and saturates at an intermediate value before decreasing into the ICB regime.  Here the parameters used were $A_t$ = 7.8 nm$^3$, $d$ = 25 \AA, $\mu$ = 7 X 10$^8$ m$^2$V$^{-1}$s$^{-1}$, T = 295 K, and C/$\alpha$ = 1.0 fF.  With these parameters the kinetic energy of an ion is approximately 1 meV, which is significantly less than the capacitive energetic barrier.  This means being in the ICB regime as expected.

To further confirm that this model captures the main
physics of this phenomenon, all-atom
MD simulations using NAMD2 were performed~\cite{Phillips_05}. In these simulations the
pores are made of silicon nitride material with a thickness of 25 \AA~in
the $\beta$-phase and have a conical shape as in figure~(\ref{fig:CB_simulation}). The bottom opening is varied
from a radius r = 5 \AA~to a radius r = 10 \AA. A given concentration of KCl
is then introduced.  With the addition of an external
constant electric field, the ionic conductance can then be probed.
Further details on the simulations can be found in work by Zwolak {\it et al.}
in References~\cite{Zwolak_09,Zwolak_10}.
%
%
\begin{figure}
\centerline{
\includegraphics[width=0.70\linewidth]{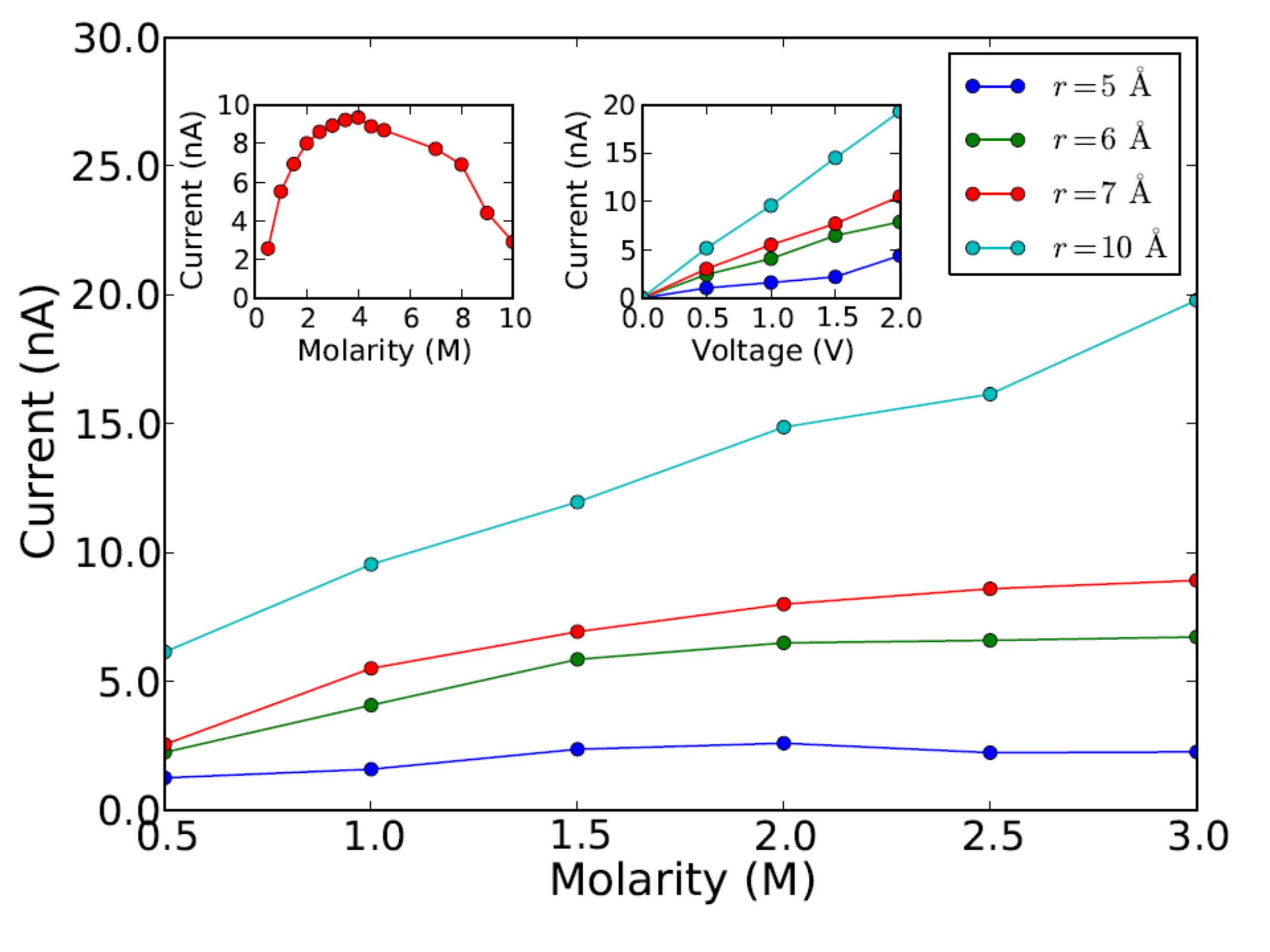}
}
\setlength{\columnwidth}{3.2in}
\caption{From Krems and Di Ventra \cite{Krems_13}: Current as a function of KCl molarity
for various neck radii and at a fixed bias of 1.0 V. For relevant
experimental molarities an almost linear increase
of the current for a radius r = 10 and a non-linear behavior
approaching a saturation at a given molarity for smaller radii is observed.
The current saturation and decrease are shown explicitly in the
left inset for a 7~\AA~neck radius pore. The other inset shows the increase
in current with voltage.  Reprinted with permission from J.\ Phys.\ Condens.\ Matter\ {\bf 25}, 065101. © IOP Publishing. Reproduced by permission of IOP Publishing. All rights reserved.
\label{fig:Ibsim} }
\end{figure}

The results of these simulations are plotted in figure~(\ref{fig:Ibsim}).
As predicted by equation~(\ref{eq:Ib2}) these calculations show an almost
linear behavior of the current as a function of bias.  For a fixed bias they show a saturation of the current as a function
of ion concentration, which is more pronounced for
pores with smaller neck radii. In the inset of figure~(\ref{fig:Ibsim}) explicit current saturation and decrease are shown for a
7 \AA~neck radius pore. Note, however, that current
saturation and decrease occur at molarities well above
the ionic precipitation limit of KCl of about 3.5 M.  These two features are therefore not predicted to be directly visible
for this configuration.

The observation of ICB should be possible in any nanopore with a constriction small enough to interfere with hydration layers, however ICB should be stronger with a V-shape pore as described above~\cite{Krems_13}.  Artificial pores of this type have been constructed~\cite{Zwolak_08}.  Experimentally, there is still some question concerning the number of surface charges on the internal walls of these pores.  However, this would only add more charge, which in turn would attract more opposite charges to accumulate inside the pore thus accelerating ICB and necessitating a smaller concentration of ions to reach current saturation.  In fact, an environment in which this surface charge could be controlled might be ideal for observing this effect.  Such control has been recently achieved experimentally by placing electrodes inside nanopores~\cite{Nam_09,Taniguchi_09,Ivanov_09}.

\section{Conclusions}

Ionic transport in nanopores is key to many cellular processes and is being explored as a method for fast and cheap DNA sequencing.  A better understanding of the electrostatics of ions in solution translocating through these pores is then of great importance to advances in this area.  This seemingly simple transport process becomes rich and complex with the inclusion of the complete electrostatic environment.  Recent work has indeed approached this problem from a microscopic viewpoint using all-atom molecular dynamics (MD) simulations to pursue a clearer understanding of ionic transport in restricted geometries.

The main description that emerges is that of ionic `quasiparticles': ions with semi-bound water molecules (hydration layers) or other ions (ionic atmosphere)~\cite{Zwolak_09,Krems_13}. Ionic atmospheres and hydration layers have been indeed known to exist for many years~\cite{Hille_01,Wright_07}.  However, their effects on ionic transport in confined geometries has been less studied. As recent experimental work on nanopores/nanochannels
has developed rapidly, it has become critical to account for these quasiparticle effects. In this review we have discussed the resulting analogies to quantum mechanics that arise in nanopore systems due to this quasiparticle behavior; namely `quantized' conductance and Coulomb blockade of ionic transport through the pore.

As stated, the hydration layers lead to quantized conductance effects in nanopore translocation~\cite{Zwolak_09}. Here, an ion with hydration layers plays the role of a classical `quasiparticle'.  If the quasiparticle
 has a spatial extent larger than the pore diameter, it will need to be broken to fit into the pore.  This breaking will accrue an energy cost thus limiting the translocation of the ion.  This phenomenon has also been confirmed using all-atom MD simulations in both nanopore and carbon nanotube transport~\cite{Beu_11b}, although its experimental
verification has not been realized yet.

Finally, a Coulomb blockade effect is predicted when taking into account the many-body interaction between ions of the same type~\cite{Krems_13}. Ionic quasiparticles of the same polarity build up in the pore up to its
capacitance, and impede transport
of extra quasiparticles with the same sign via electrostatic repulsion. This accumulation then leads to a non-ohmic conductance as a function of molarity which is different from that observed under other  conditions~\cite{Wright_07,Powell_07,Cruz-Chu_09,Coronado_80,Lu_94,Rostovtseva_98}. One difference from the quantum-mechanical Coulomb blockade effect is that the ions in solution are not in the quantum regime and therefore do not need to satisfy exclusion
statistics~\cite{Krems_13}. As technology develops, it is becoming possible to create pores comparable in size to the hydration layers~\cite{Nam_09,Taniguchi_09,Ivanov_09}. Therefore, we expect
that further experimental progress should enable a clearer observation of these effects and their role in many emerging technologies.

\label{sec:Conclusions}

\section{Acknowledgments}

This work was supported in part by the National Human Genome Research Institute of NIH. We
thank J. Lagerqvist, M. Krems, J. Wilson, and M. Zwolak for their
collaboration on many of the projects reviewed here.\\



\end{document}